\documentclass[aps, prb, 10pt ,a4paper,twocolumn,floatfix,showpacs,showkeys,amsmath,amssymb,nobibnotes,altaffilletter]{revtex4-1}
\usepackage{graphicx}
\usepackage{epstopdf}
\usepackage{color}
\usepackage{hyperref}

\begin{document}

\title{S-shaped current--voltage characteristics of organic solar devices}

\author{A. Wagenpfahl$^1$}
\author{D. Rauh$^2$}
\author{M. Binder$^1$}
\author{C. Deibel$^1$}
\author{V. Dyakonov$^{1,2}$}
\affiliation{$^1$Experimental Physics VI, Julius-Maximilians-University of W{\"u}rzburg, 97074 W{\"u}rzburg, Germany.\\$^2$Bavarian Centre for Applied Energy Research (ZAE Bayern), 97074 W{\"u}rzburg, Germany.}

\date{\today}

\begin{abstract}

Measuring the current--voltage characteristic of organic bulk heterojunction solar devices sometimes reveals an s-shaped deformation. We qualitatively produce this behaviour by a numerical device simulation assuming a reduced surface recombination. Furthermore we show how to experimentally create these double diodes by applying an oxygen plasma etch on the indium tin oxide (ITO) anode. Restricted charge transport over material interfaces accumulates space charges and therefore creates s-shaped deformations. Finally we discuss the consequences of our findings for the open circuit voltage $V_{oc}$.\\[2ex]
{\bf \href{http://dx.doi.org/10.1103/PhysRevB.82.115306}{Phys.\ Rev.\ B. 82, 115306 (2010)}}

\end{abstract}



\maketitle
\newpage

\section{Introduction}

Since the efforts to build organic photovoltaic devices began, groups around the world tried to create the ideal semiconductors and device configurations.\cite{brabec2008book, deibel2010review3} The first experiments with bilayer solar cells where soon superseded by devices with a blended active layers of two different semiconductors. Those bulk-heterojunction solar cells nowadays achieve power conversion efficiencies of up to 8~\% bringing the economical break even within reach.\cite{liang2010, green2010}

Experiments with innovative organic semiconductors, but also well-known systems, sometimes show an s-shaped deformation of the device current--voltage characteristic. Instead of an exponential current--voltage ratio as expected  for diode structures, the response of such a device in conducting direction shows a local saturation and a later again increasing current around at a certain applied voltage region. A similar behavior was reported for inorganic copper indium gallium diselenide (CIGS) based devices which was attributed to the influence of a chargeable cadmium sulfide buffer layer, creating an illumination dependent energetic barrier.\cite{eisgruber1998} In organic multilayer solar cells this deformation was observed several times in literature and was attributed to surface dipoles.\cite{kumar2009, uhrich2007, schulze2006}

For numeric simulations a careful description of device interfaces therefore should lead to a double diode characteristic. In most publications metal-organic interfaces are described by a Schottky contact. The boundary condition for the charge carrier densities therefore is defined by the thermionic emission theory resulting in a fixed charge carrier density at the interface.\cite{selberherr1984, koster2005b, hausermann2009} Restrictions in the ability to transfer charges from one side of the junction to the other can be expressed by the surface recombination rate.\cite{grove1967} By the latter, models such as mirror charge effects at metal--organic interfaces can be described, resulting in surface charge carrier densities different than predicted by the thermionic emission theory.\cite{scott1999a} For organic solar cells the importance of low surface recombination rates on the power conversion performance was already discussed, mainly for minority charges.\cite{kirchartz2008, wagenpfahl2010b} 

Beside these boundary conditions at the interfaces various models for physical effects inside the solar cell bulks are available for numerical simulations. Optical interference calculations are used to determine the spatial charge carrier generation by absorbed light for different device geometries.\cite{sievers2006, pettersson1999} Assuming the existence of charge transfer states during the generation process an electric field dependence can be assigned based on the assumption of Gaussian distributed exciton binding distances.\cite{koster2005b} From the energetic and spatial disorder in organic materials charge transport models for the mobility of electrons and holes could be deduced using Monte Carlo simulation techniques.\cite{bassler1993, novikov1998, tessler2009} Combinations of those models are often used to compare and evaluate experimental data.\cite{lacic2005, buxton2007} Nevertheless none of these models is able to predict s-shaped current--voltage characteristics in simple device geometries as discussed in this paper.
 
Within this article, we will show how to experimentally create s-shaped current--voltage characteristics of organic bulk heterojunction solar cells by an extended oxygen plasma etching of the indium tin oxide (ITO) anode. Using a numerical simulation, we qualitatively reproduce this double diode assuming reduced surface recombination velocities at the anode. We will discuss the effects able to create s-shaped deformations as well as the consequences of our findings for organic photovoltaic devices.

\section{Methods}\label{sec:methods}

\subsection{Experimental methods}

For the solar cell preparation structured ITO glass was successively cleaned in soap water, aceton and isopropanol in each case for 10 min in an ultrasonic bath. The anode was exposed to an oxygen plasma etching with a certain nitrogen concentration for various time scales to change the ITO work function.\cite{kim1998} On this substrate, an optional layer of poly (3,4-ethylenedioxythiophene) poly(styrenesulfonate) (PEDOT:PSS) was spincoated and transfered into a nitrogen atmosphere before annealing the PEDOT:PSS for 10 minutes at 130 $^{\circ}$C. The active layer consisting of poly (3-hexyl thiophene) (P3HT) as donor and [6,6]-phenyl-C61 butyric acid methyl ester (PCBM) as acceptor was spin coated from a chlorobenzene solution (ratio 1:0.7). Finally, a second annealing step at 130 $^{\circ}$C for 10 minutes and the evaporation of the metal Ca/Al cathode completed the solar cell assembly. The materials were purchased from H.C. Starck (PEDOT:PSS), Rieke Metals (P3HT) and Solenne (PCBM). 
The current--voltage characteristics were recorded by a Keithley 237 source measurement unit under an artificial AM1.5g sun spectrum with a spectral radiance of 100 $\text{mW/cm}^2$, simulated by a 300W xenon lamp adjusted by a mismatch factor. Layer thicknesses and the ITO roughness were determined by a Veeco Dektak 150 profilometer.

\subsection{Numerical methods}

In order to simulate current--voltage characteristics, we use the elliptic differential equation system of Poisson equation,
\begin{equation}
	\frac{\partial F}{\partial x} = \frac{q}{\epsilon_0 \epsilon_r} (p(x) - n(x)) \label{eqn:poisson},
\end{equation}
two charge transport equations for electrons and holes, 
\begin{eqnarray}
	J_n &=& q(n \mu_n F + D_n \partial n / \partial x)\label{eqn:transportn}\\
	J_p &=& q(p \mu_p F - D_n \partial p / \partial x),\label{eqn:transportp}
\end{eqnarray}
and the continuity equations in steady state,
\begin{eqnarray}
	\frac{\partial J_n}{\partial x} &=& q \left( G(x) - R(x)  \right)\label{eqn:continuityn}\\
	\frac{\partial J_p}{\partial x} &=& -q \left( G(x) - R(x)  \right).\label{eqn:continuityp}
\end{eqnarray}	

Discretized and combined with models for charge carrier generation $G(x)$ and recombination $R(x)$, this can be solved numerically using approximations according to Gummel and Scharfetter.\cite{gummel1964, scharfetter1969, deibel2008a, selberherr1984}

As basic parameters information on the charge carrier mobilities for electrons $\mu_n$ and holes $\mu_p$, the corresponding effective charge carrier densities $N_c$, $N_v$, the temperature $T$ and the material effective dielectric permeability $\epsilon_r$ are required. $F$ denotes the electical field.

Inside the blend, excitons generated by light on the polymer chains are splitted up on a femtosecond timescale when reaching a donor-acceptor (polymer-fullerene) interface.\cite{sariciftci1992} The hole remains on the donor's highest occupied molecular orbital (HOMO) whereas the electron is transferred to the lowest unoccupied molecular orbital (LUMO) of the acceptor due to the energetic advantage of this state. Assuming an effective medium approach this process is allowed everywhere inside the active layer leading to free charge carriers which are transported through the device. The energy difference between the acceptor's LUMO and the donnor's HOMO energy is used as the effective band gap $E_g$. For all presented calculations, the charge carrier generation rate $G(x)~=~G_0$ is set to be constant throughout the device.

The charge carrier recombination process inside the active layer which is bimolecular and non-geminate is described by the Langevin theory\cite{langevin1903, deibel2009}
\begin{equation}
R = \frac{q}{\epsilon_0 \epsilon_r}\left( \mu_n + \mu_p \right) \left(np - n_i^2 \right).
\label{eqn:langevin}
\end{equation}
This rate is proportional to the product of charge carrier densities for electrons $n$ and holes $p$ until the thermally activated ground state is reached, which is indicated by the intrinsic charge carrier density
\begin{equation}
n_i = \sqrt{N_c N_v} \exp \left(E_g/2k_BT \right), 
\end{equation}
using the Boltzmann constant $k_B$ and the dielectric vacuum permeability $\epsilon_0$.

According to the thermionic emission theory, we calculate the thermally activated charge carrier densities at the contacts by the energetic distance between the anode work function and hole conducting HOMO level at the interface, defined as the injection barrier $\Phi_p$
\begin{eqnarray}   
	n_{th} &=& N_c \exp \left( - E_g + \Phi_p ) / k_B T \right)\label{eqn:inj_barriers_n}\\		 
	p_{th} &=& N_v \exp \left( -\Phi_p / k_B T \right). \label{eqn:inj_barriers_p}
\end{eqnarray}
Vice versa the injection barrier $\Phi_n$ is the difference between the electron conducting LUMO and the cathode Fermi level at the cathode.  

In analogy, the local charge carrier density inside the active layer can be calculated from the energetic distance of conductive bands and their corresponding quasi-Fermi levels $E_{Fn}$, $E_{Fp}$,
\begin{eqnarray}
	n &=& N_c \exp( (E_{Fn} - E_{LUMO})/k_BT)\label{eqn:n}\\
	p &=& N_v \exp( (E_{HOMO} - E_{Fp})/k_BT).\label{eqn:p}
\end{eqnarray}

In order to represent metal--semiconductor charge transfer processes in a general way, we use the surface recombination rate for electrons and holes defined by
\begin{equation}
	J_{s} = q S \left(n - n_{th} \right).
	\label{eqn:surf}
\end{equation}
Since this equation does not exclusively describe recombination, but an extraction of charge carriers out of the bulk it also can be considered as an extraction rate. In contrast to the thermionic emission theory the amount of surface charges $n$ is changed by the surface recombination velocity $S$, which defines the current transferred accross the metal--semiconductor interface. As we consider an active material between two metallic electrodes, we assume four independent surface recombination velocities $S_{n,p}^{a, c}$, for electrons and holes at each electrode, as listed in Table~\ref{tab:param}. Due to the different amount of charge carriers, holes and electrons at the anode are denoted as majority and minority charge carrieres. Vice versa at the cathode, holes and electrons are the minorities and majorities. Straightforward we distinguish between majority and minority surface recombination velocities.

All presented calculations were performed with the parameter set given in Table \ref{tab:param} unless otherwise noted.

\begin{table}[h]
	\centering
		\begin{tabular}{lll}
			\hline
			\hline
			parameter & value & description \\
			\hline
			$E_{g}$	& $1.1~\text{eV}$ & effective band gap\cite{vandewal2008, veldman2009} \\
			$\Phi_n$, $\Phi_p$ & $0.0$ eV, $0.1~\text{eV}$ & injection barriers\\
			$\mu_n,~\mu_p$ & $10^{-8}~\text{m}^2\text{/Vs}$ & mobilities\cite{baumann2008}\\
			$L$ & $100~\text{nm}$ & active layer thickness\\
			$G_0$ & $6.0\cdot10^{27}~\text{m}^{-3}\text{s}^{-1}$ & generation rate\\
			$T$ & $300~\text{K}$ & temperature\\
			$N_{c},~N_v$ & $ 10^{26}~\text{m}^{-3} $ & effective density of states\\		
			$\epsilon_r$ & $3.4$ & relative static permittivity\cite{persson2005}\\	
			\\
			\hline
			\hline
			\multicolumn{3}{l}{surface recombination velocities:}\\
			\hline
			$S_p^{c}$ & $\rightarrow\infty$ & minority, cathode\\
			$S_n^{a}$ & $\rightarrow\infty$ & minority, anode\\
			$S_n^{c}$ & $\rightarrow\infty$ & majority, cathode\\			
			$S_p^{a}$ & $10^{-9}$~m/s & majority, anode\\
			\hline
			\hline
		\end{tabular}
	\caption{Standard parameters used in the macroscopic simulation except as noted otherwise. An infinite surface recombination velocity is represented as $10^{50}\text{m/s}$ in the numeric calculations.}
	\label{tab:param}
\end{table}

\section{Results}

\subsection{Experimental results}
\begin{figure}
   \centering
   \includegraphics[width=9.0cm]{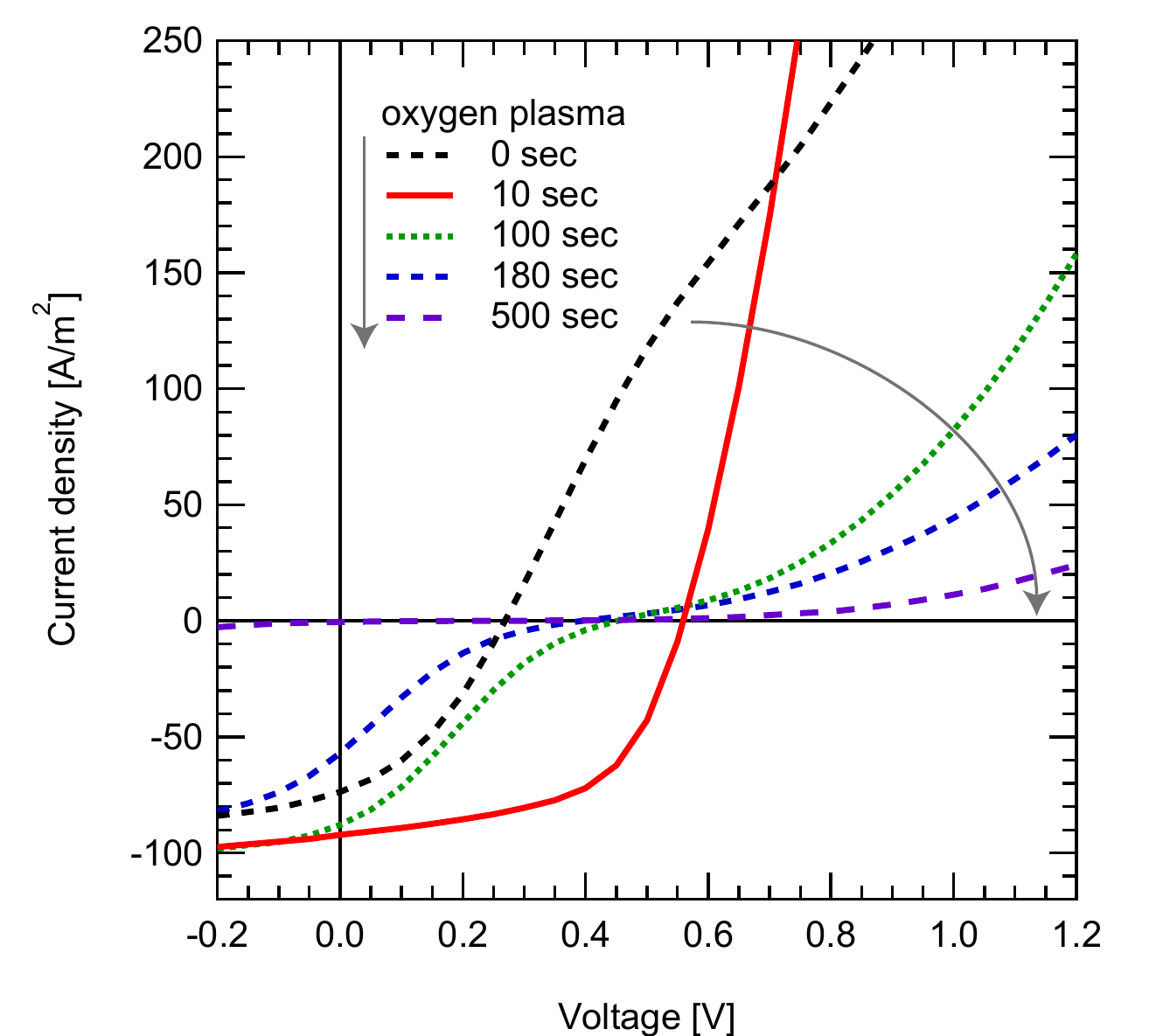} 
   \caption{(Color online) Measured current--voltage characteristics of a bulk heterojunction solar cells without PEDOT:PSS layer under illumination. The cell with an additional 10 second oxygen plasma treatment of the ITO substrate shows a regular solar cell behavior whereas an extension leads to a distinct s-shaped behavior reducing the solar cell efficiency.}
   \label{fig:ivexp}
\end{figure}

In Figure \ref{fig:ivexp} typical current--voltage characteristics of the P3HT:PCBM organic solar cells without a PEDOT:PSS layer under various oxygen plasma treatment timescales are shown. Within the first 10 seconds the power conversion efficiency increases from 0.72~\% to 2.9~\%. The open circuit voltage is enhanced from 270~mV to 560~mV, the fill factor from 0.36 to 0.56 and the short circuit current from 7.37~mA/cm$^2$ to 9.23~mA/cm$^2$. 

After 500 seconds of oxygen plasma treated the samples undergo an efficiency loss to 0.0017~\%. This can be attribute to the drastically drop of short circuit current to 0.047~$\text{mA/cm}^2$. The open circuit voltage decreases slightly to 235~mV, the fill factor to 0.15. The meassured roughness of the ITO substrate was not changed during the oxygen plasma etching within the profilometer noise level of 5~nm. The active layer thickness of all measured samples was about 230~nm. 

Deposing a PEDOT:PSS layer on top of the treated ITO anode slightly improves the initial solar cell working efficiency. In contrast to a cell without PEDOT:PSS the first ten seconds of oxygen plasma etch do not improve this value. Longer treatment durations result in an equivalent s-shaped characteristics behavior but with a constant $V_{oc}$ level (not shown).

\subsection{Numerical results}
\begin{figure}
	\centering\includegraphics[width=9.0cm]{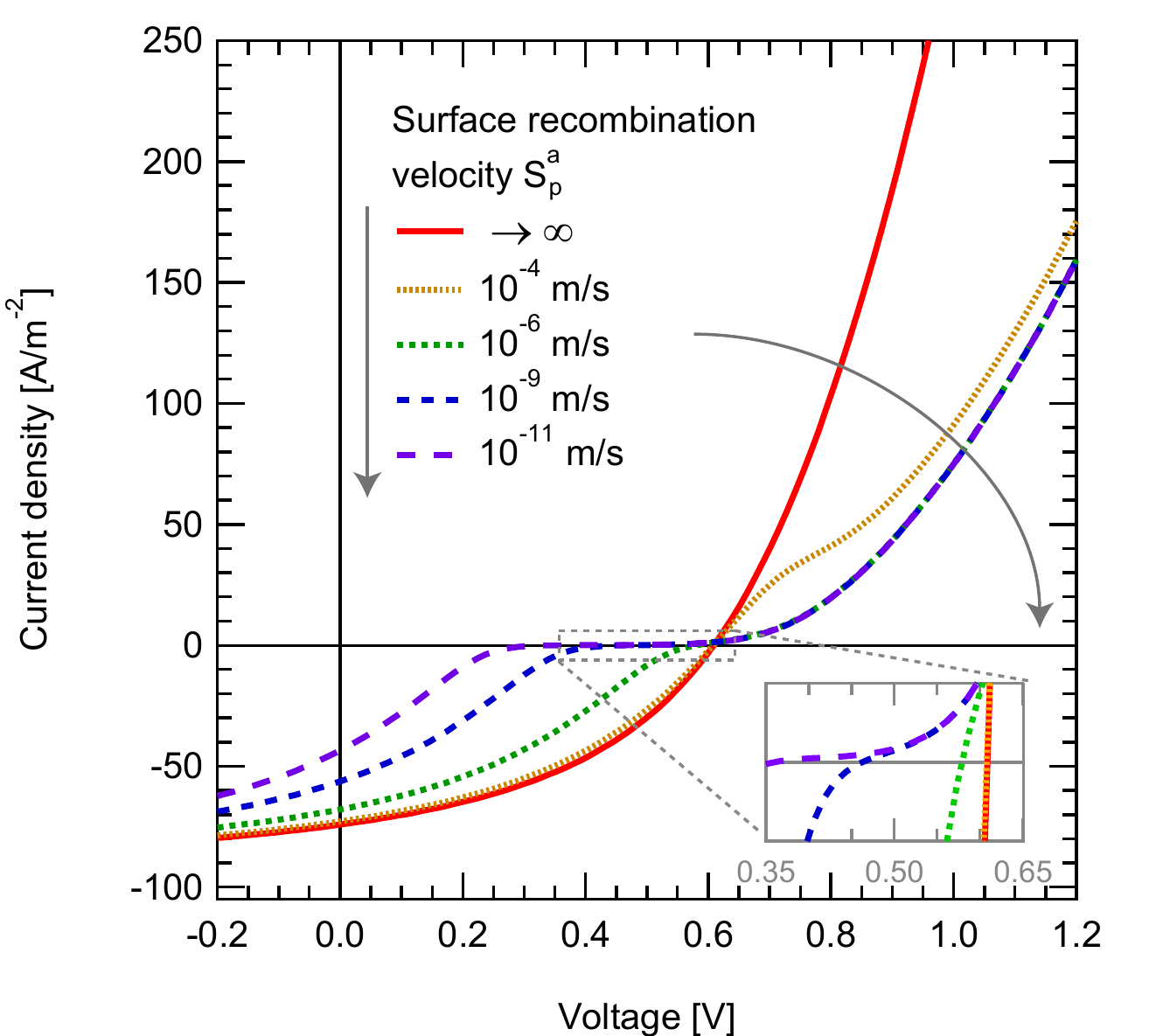}
	\caption{(Color online) Simulated s-shaped current--voltage characteristics due to a reduced majority surface recombination velocity $S_p^a$ for holes at the hole conducting anode. With decreasing velocity, the s-shape gets more distinctive limiting the power converion efficiency by diminishing open circuit voltage, fill factor and short circuit current. At the lower right corner a magnification of the $V_{oc}$ dependence on decreasing surface recombination velocity is shown.}
	\label{fig:IVlinear}
\end{figure}

In order to explain the experiment qualitatively, we focus on the effect of metal--semiconductor interfaces, namely a reduction of the surface recombination velocities at the hole extracting contact (anode) for majority charge carriers. By limiting the majority surface velocity from infinity (numerically $\approx 10^{50}$~m/s) to $10^{-4}$~m/s, $10^{-6}$~m/s, $10^{-9}$~m/s and $10^{-11}$~m/s, s-shaped IV-curves are produced as shown in Figure~\ref{fig:IVlinear}. With the parameters given in Tab.~\ref{tab:param} and infinite surface recombination velocities, we retrieve an open circuit voltage of 609~mV. For the given reduced surface recombination velocities, $V_{oc}$ drops down to 360~mV. The calculated short circuit current decreases from 7.41~$\text{mA/cm}^2$ to 4.34~$\text{mA/cm}^2$, the fill factor from 0.41 to 0.18.

\section{Discussion} 

Metal--organic junctions posses an intrinsic charge carrier density described by the injection barrier according to the thermionic emission theory. By this parameter no prediction on the amount of conducted charge is made. If there are any restrictions on the charge extraction process this needs to be expressed by the surface recombination velocities according to Eqn.~(\ref{eqn:surf}). If we assume a constant current that has to be transported through the interface, a reduction of the surface recombination velocity $S$ has to be compensated by additional charges at the junction ($n - n_{th}$), maintaining the passed currrent. These piled-up charges therefore create an applied voltage dependent space charge region which modifies the device energetic band structure. In Figure~\ref{fig:IVlinear} the effect of constant majority surface recombination velocities at the anode is shown creating a typical s-shaped current--voltage curve.

\begin{figure}
	\centering\includegraphics[width=9.0cm]{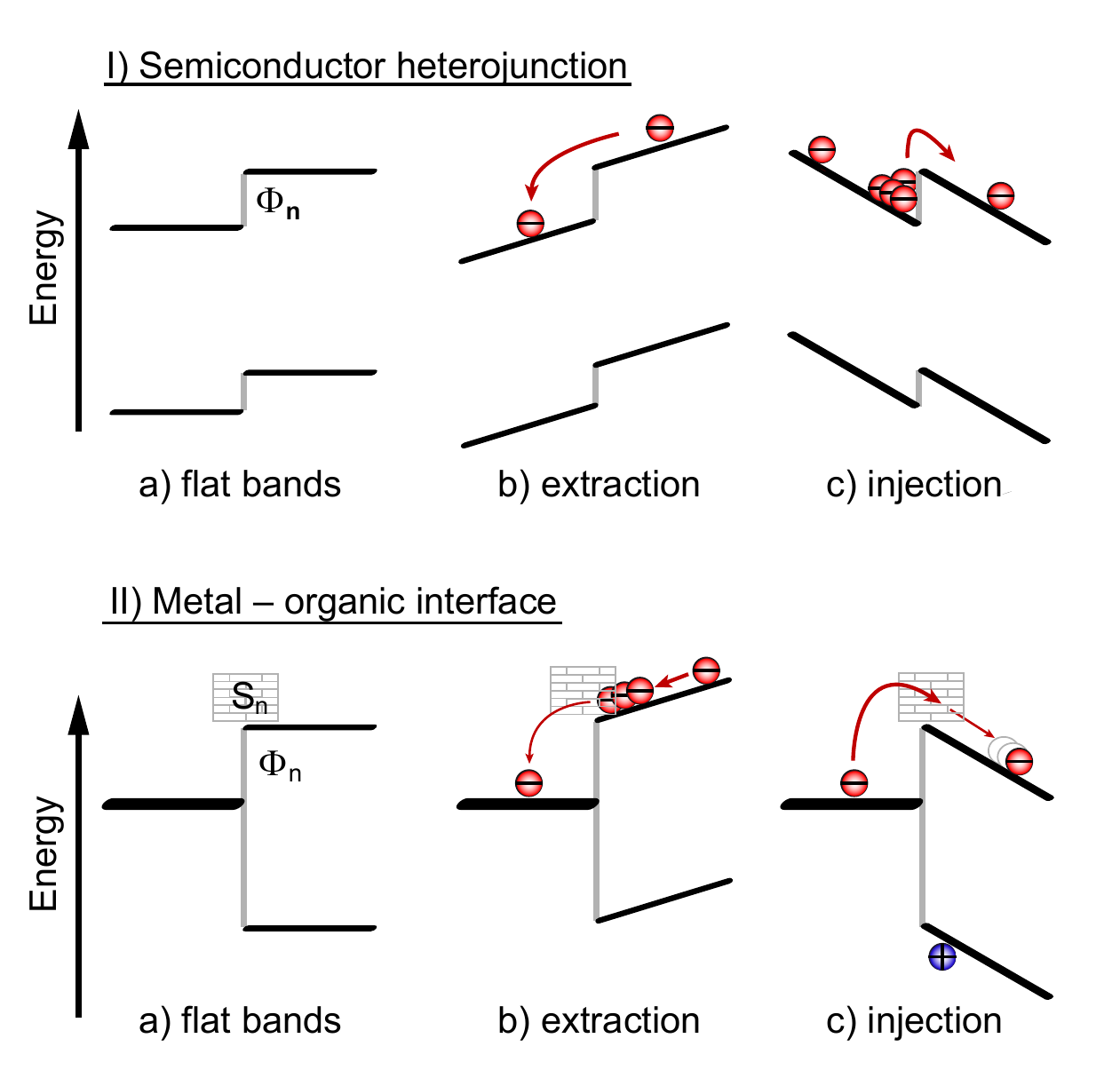}
	\caption{(Color online) Space charge accumulation in semiconductor devices. (Upper row, a) At a semiconductor heterojunction the two LUMO energy levels possess a work function difference of $\Phi_n$ at the interface. (b) Additional electrons above the intrinsic concentration, injected by an external electric field cross this interface from an electric unfavorable state to the more favorable one without restrictions. (c) If electrons are forced to overcome this energetic barrier in the opposite direction, a local potential well is established creating a pile up of electrons at the heterojunction.
(Lower row, a) At a metal--organic interface the work function difference between the metal Fermi level and the semiconductor LUMO level $\Phi_n$ defines the intrinsic electron concentration at the interface. Due to the constant metal work function no potential well or space charge can be created. (b) By a finite surface recombination velocity $S_n$ electron transport through the interface is limited. Electrons which are transported faster towards the interface than they are removed create a space charge. (c) The amount of injected charge carriers is reduced by finite surface recombination velocities creating a local charge carrier depletion zone at the interface.}
	\label{fig:scheme}
\end{figure}

From literature we known planar semiconductor heterojunctions to cause equivalent s-shaped characteristics. The physical origin of this effect is depicted in the upper row of Figure~\ref{fig:scheme}. The heterojunction is defined (I a) by the energy differences between the two LUMO ($\Phi_n)$ and the two HOMO bands ($\Phi_p$) of both semiconductors directly at the interface. Driven by an electric field (I b) additional electrons above the intrinsic neutrality level can cross such an interface without transport problem if they are transfered from the energetically unfavorable level to a more favorable one. However, by band bending of the two semiconductors a local potential well is created if the polarity of the applied voltage is changed (I c), charging a depending local space charge and creating an s-shaped current--voltage characteristic.\cite{schulze2006, uhrich2007, uhrich2008} The same effect has been seen in copper indium gallium diselenide (CIGS) devices where such a s-shape behavior was assigned to by light filled traps in an intermediate cadmium sulfide layer creating an illumination dependent energetic barrier.\cite{eisgruber1998}

In device geometries such as bulk heterojunctions, without any planar semiconductor heterojunction, this approach cannot be applied. Due to the constant metallic work function metals are not able to create potential wells as shown in Figure~\ref{fig:scheme} (II). The injection barrier $\Phi_n$ (II a) defines the intrinsic charge carrier density at the interface as defined by the Boltzmann equation (Eqn. \ref{eqn:inj_barriers_n}). Hence creating a local space charge needs the definition of an additional parameter, the surface recombination velocity. If, under extraction conditions (II b) charges are transported faster towards the interface than they can be extracted, those charges will pile up creating a space charge depending upon the applied electric field. Under injection conditions (II c) the injection of charges gets reduced if the surface recombination velocity is expected to be constant. Hereby the influence of the intrinsic holes rises, creating a local depletion zone. The injection barrier at metal--organic interfaces is not able to create s-shapes on its own.

As explained in Figure~\ref{fig:scheme} for electrons the issue holds for holes as well. For moderatly reduced values of $S^a_p$ ($10^{-4}$~m/s) (Figure~\ref{fig:IVlinear}), an \mbox{s-shaped} current--voltage characteristic is created at higher injecting voltages. It does not influence the solar cell efficiency parameters $V_{oc}$, $J_{sc}$ and the fill factor significantly. Decreasing the surface velocity further, the deformation will approach $V_{oc}$ creating a horizontal current plateau. Hence, all mentioned solar cell parameters are reduced.

\subsection{Energetic structure}

\begin{figure}
	\centering\includegraphics[width=9.0cm]{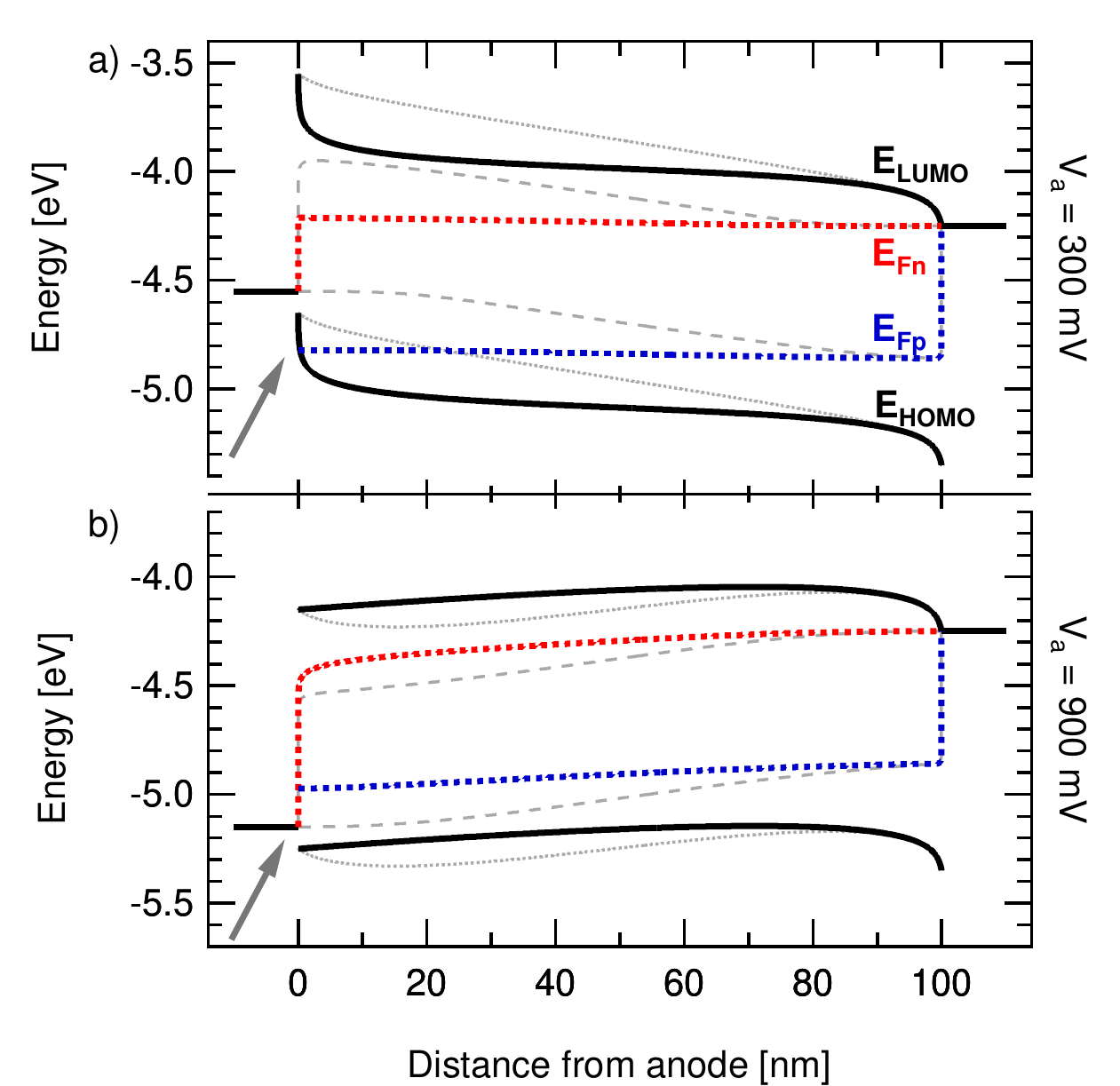}
	\caption{(Color online) Energy structure of a solar cell with the energetic levels of HOMO and LUMO (solid and dotted lines) as well as their corresponding quasi-Fermi levels $E_{Fn}$ and $E_{Fp}$ (dashed red and blue lines) under illumination. The thick lines indicate the structure under a surface recombination velocity of $S_p^a=10^{-9}$~m/s. Thin lines correspond to perfectly conducing contacts ($S_p^a\rightarrow \infty$). In the case of hole injection by the anode (compare to Fig. \ref{fig:scheme} (II b)) at an applied voltage of 300~mV the quasi-Fermi level of holes split up at the anode. Describing the charge carrier density this generates a space charge combined with a higher electric field ($\propto \partial E_{LUMO} / \partial x$). At a voltage of 900~mV holes are injected into the bulk by the anode (compare to Fig. \ref{fig:scheme} (II c)). Created by the surface recombination, less majority charges are present at the interface creating an electron depletion zone which decreases the local electric field. The cathode energy level (right side) was set to a fixed value of -4.25~eV.}
	\label{fig:band_app}
\end{figure}

The deformation of the current--voltage characteristics originates from the internal energetic band structure and charge carrier distributions of electrons and holes. Figure~\ref{fig:band_app} shows the device band structure under applied voltages of (a) $V_a = 300$~mV and (b) $V_a = 900$~mV, hence under extracting and injecting conditions, respectively. The energy levels of LUMO and HOMO bands ($E_{LUMO}$, $E_{HOMO}$) are indicated by the thick solid lines, those of quasi-Fermi levels $E_{Fn}$, $E_{Fp}$ by the blue and red dashed lines. At the anode (left contact) a limiting surface recombination velocity of $S^a_p = 10^{-9}$~m/s is assumed. For comparison, the underlying gray lines show the same solar cell with an infinite surface recombination velocity which will not lead to an s-shape deformation.

From these diagrams, position dependent information on the electric field, the charge carrier density and the current flux inside the device can be determined. The electric field is proportional to the slope of the energetic bands $E_{LUMO}$ and $E_{HOMO}$ ($F = - q \partial E / \partial x $). The local charge carrier densities are exponential functions of the energetic distance between the band energy level and the corresponding quasi-Fermi level (Eqns. (\ref{eqn:n}), (\ref{eqn:p})). Multiplied with the slope of the related quasi-Fermi level these charge carrier densities result in the local current densities ($J_n = \mu_n n \partial E_{Fn}/ \partial x$) caused by drift as well as diffusion.\cite{sze2007}

Within the solar cell working regime charges generated by light are extracted from the active layer into the metal electrodes. A reduced majority surface recombination velocity $S^a_p$ will hinder an efficient holes extraction by the anode and holes are piled up at the interface (Figure~\ref{fig:band_app} (a)). Consequently, at the anode interface the quasi-Fermi level of holes is shifted towards its HOMO level creating a quasi-Fermi level discontinuity between $E_{Fp}$ and the anode Fermi level. The interface hole density is higher than predicted by the thermionic emission theory. Additionally to the already high hole concentration at the anode interface for infinite surface recombination velocities additional majority charges will be added creating a space charge. Due to Coulombic interaction these charges increase the local electric field at the contact and consequently reduce the potential drop inside the semiconductor bulk.\cite{mihailetchi2003a} In a solar cell dominated by drift, charges are transported more slowly through the device, thus increasing the probability of recombination.

At voltages above the built-in potential additional charges will be injected into the semiconductor bulk by the metal contacts (Figure~\ref{fig:band_app}(b)). A finite majority surface recombination velocity therefore will reduce the number of injected holes at the anode interface. Thus, the quasi-Fermi level for holes $E_{Fp}$ at the interface is shifted away from $E_{HOMO}$. Within the interface region fewer holes and more electrons (due to the limited bulk recombination rate, Eqn.~(\ref{eqn:langevin})) will be available. Consequently the electric field at the interface is reduced and --- as in case of Figure~\ref{fig:band_app}(b) --- is even able to change signs. The transition between both cases creates an s-shaped kink in the current--voltage characteristics of organic solar cells and leads to a space charge limited current at higher voltages.

\subsection{Open circuit voltage}

\begin{figure}
	\centering\includegraphics[width=9.0cm]{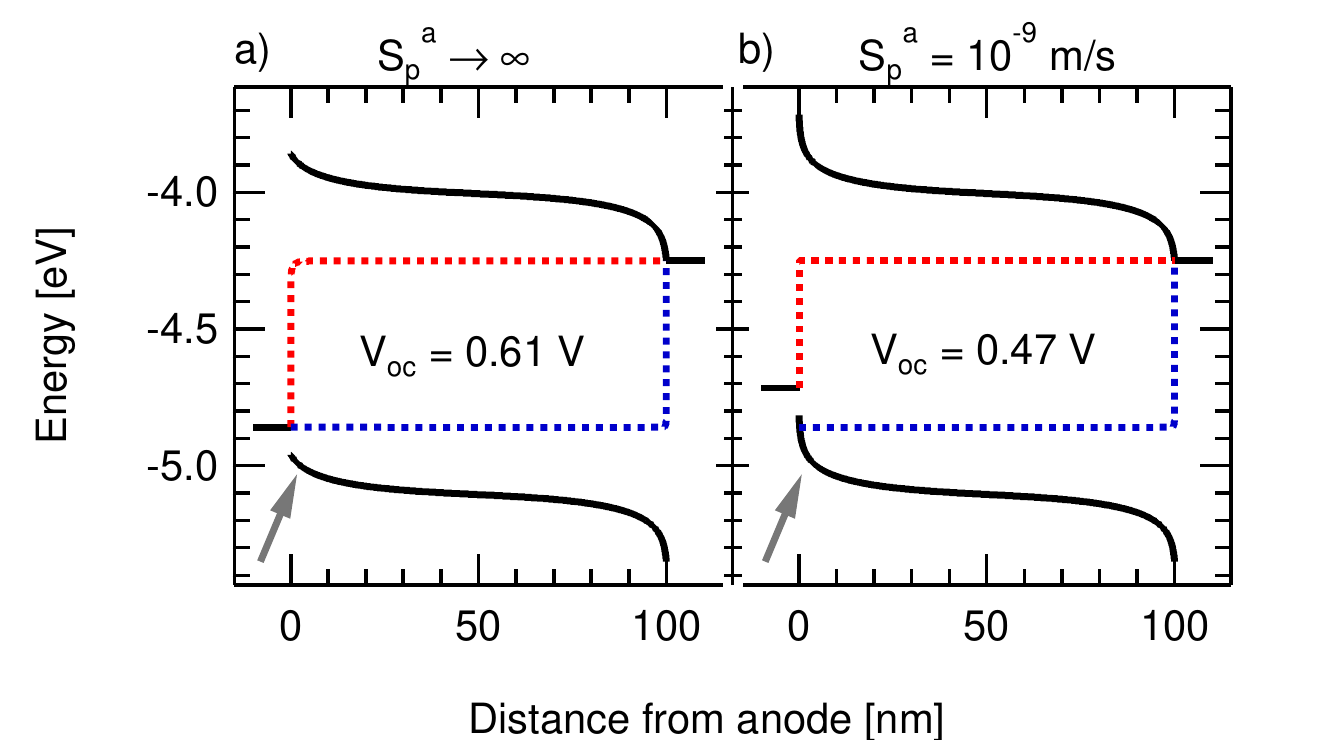}
	\caption{(Color online) Device band diagram at the open circuit voltage for surface recombination rates of (a) infinity and (b) $S_p^a=10^{-9}$~m/s. Due to the condition of zero net current ($\vec{J} = \vec{j_n} + \vec{j_p} = 0$) holes pile up at the anode creating a space charge which directly reduces $V_{oc}$. Whereas the quasi-Fermi distribution changes only slightly due to bulk recombination effects, but splits up directly at the contact the band bending and hence the electric field at the anode increases.}
	\label{fig:band_voc}
\end{figure}

With decreasing surface recombination velocity the open circuit voltage drops according to Figure~\ref{fig:IVlinear}. In Figure~\ref{fig:band_voc} the band diagrams for (a) infinite and (b) finite ($S^a_p = 10^{-9}$~m/s) surface recombination velocities with their corresponding open circuit voltages of 0.61 V and 0.47 V are shown. As in the case of charge carrier extraction the electric field at the anode is enhanced. Nevertheless the quasi-Fermi level distributions change only slightly except for their split-up directly at the anode contact.

Under open circuit conditions, the total current at every location inside the device must be zero. Accordingly, the internal current densities for electrons and holes are equal and of different sign, $\vec{J}(x) = \vec{j_n}(x) + \vec{j_p}(x) = 0$. Since an unlimited (infinite) minority surface recombination always creates a steady diffusive recombination current towards the anode, this current has to be neutralized by an equivalent majority surface recombination current. If the surface recombination velocity according to Eq.~(\ref{eqn:surf}) is reduced, the condition of zero net current can only be fullfilled by an accumulation of majority charges at the surface. The resulting creation of a space charge leads to an s-shape around $V_{oc}$, and can even reduce $V_{oc}$ (Figure~\ref{fig:band_voc}).

\begin{figure}
	\centering\includegraphics[width=9.0cm]{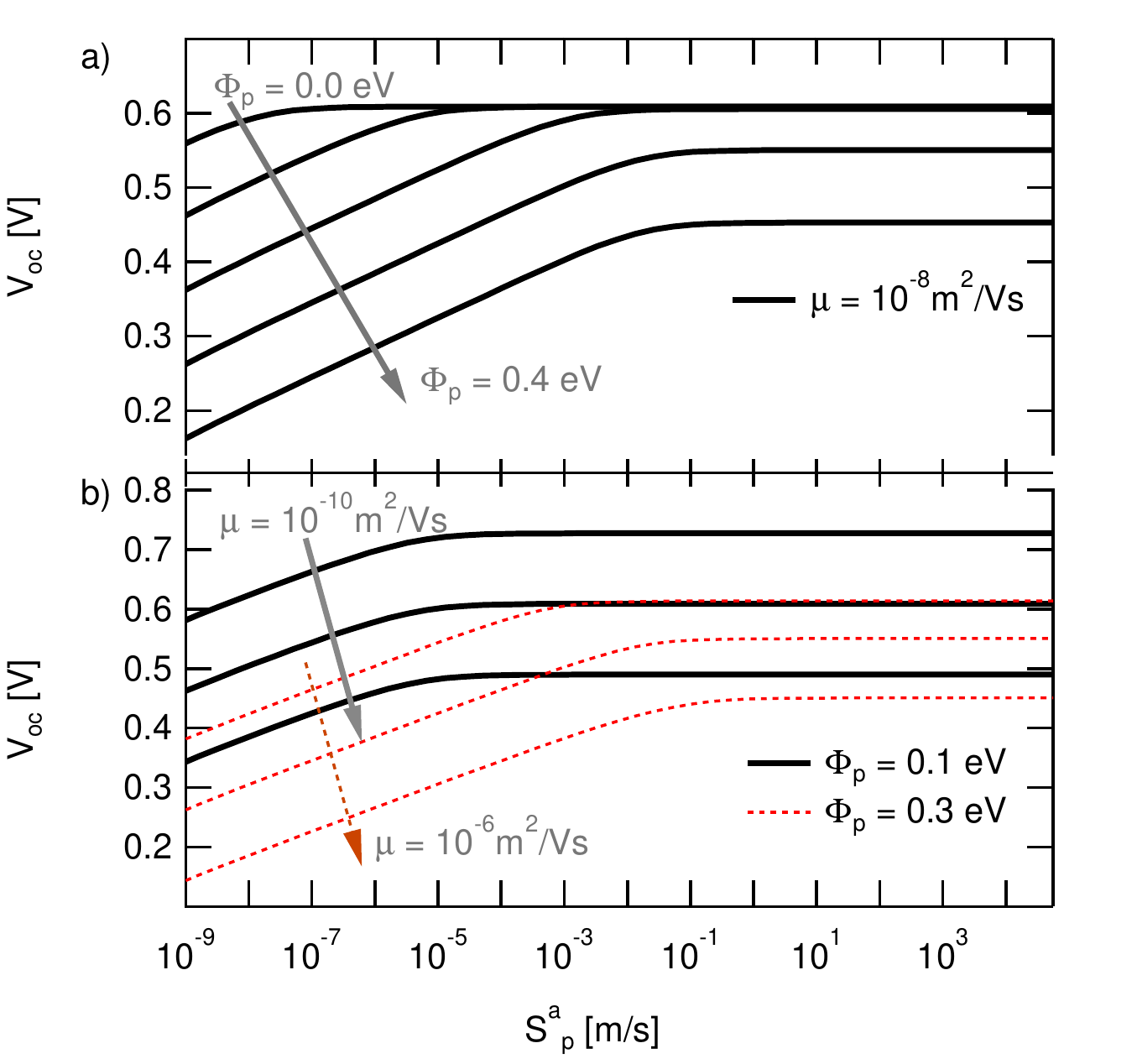}
	\caption{(Color online) Open circuit potential plotted versus the hole surface recombination velocity at the anode. Assuming balanced charge carrier mobilities ($\mu_n = \mu_p$) (a) shows a variation of the anode injection barrier from 0.0 to 0.4 eV in steps of 0.1eV whereas (b) presents the same solar cell under constant injection barriers and charge carrier mobilities of $10^{-10}$~m$^2$/Vs, $10^{-8}$ m$^2$/Vs and $10^{-6}$~m$^2$/Vs. At high surface recombination velocities no dependence of $V_{oc}$ from the surface recombination velocity can be observed in both cases. A reduction below a critical point results in a linear decreasing of $V_{oc}$ on the semi logarithmic scale which can be attributed to a logarithmic dependence of the local electric field at the contact on the majority charge carrier density.}
	\label{fig:Voc}
\end{figure}

\subsection{Analytical approximation of $V_{oc}$}
Considering the dependence of $V_{oc}$ on the anode majority recombination velocity (Figure~\ref{fig:Voc} a) two different regimes can be distinguished. At high majority surface recombination velocities $V_{oc}$ remains constant whereas at lower values $V_{oc}$ decreases logarithmically. Additionaly the open circuit voltage is lowered by the injection barrier $\Phi_p$.

The open circuit voltage can be calculated, using the local electric field $F(x) =  \partial / \partial x \left(q E_{HOMO}(x)\right)$ by\cite{cheyns2008}
\begin{equation} 
	V_{oc} = \frac{E_g - \Phi_n - \Phi_p}{q} - \int_0^L F(x) \text{d}x.
	\label{eqn:Voc}
\end{equation}

At high surface recombination velocities, $V_{oc}$ is dominated by the injection barriers. The barrier $\Phi_p$ changes the thermal equillibrium charge carrier densities $n_{th}$, $p_{th}$ (Eqns.~(\ref{eqn:inj_barriers_n}), (\ref{eqn:inj_barriers_p})) to balanced conditions at the contact. Since the electric field generated by charge carriers is smaller for balanced densities, the band bending at the interface is reduced. Potential losses due to injection barriers are compensated, hardly reducing the effective open circuit voltage. If the band bending at the anode is fully compensated, any higher injection barrier therefore linearly reduces $V_{oc}$.

With decreasing $S^a_p$, additional majority charges are piled up at the anode. This enhances the electric field at the interface. In order to be able to influence $V_{oc}$, the additional space charge must be significant compared to the electric field of infinite surface recombination solar cell contacts, a minimal and therefore critical additional charge carrier density is required, indicated by a kink on the $V_{oc}(S^a_p)$ diagram.

In order to understand the slope of $V_{oc}$ on lower recombination velocities, the electric field at the contact can be calculated, solving the Poisson equation in one dimension (Eqn.~(\ref{eqn:poisson})) together with the current transport equations for electrons (Eqn.~(\ref{eqn:transportn})) and holes (Eqn.~(\ref{eqn:transportp})) for the electric field. Under the condition of zero net current at $V_{oc}$, we retrieve
\begin{equation}
	F(x) = \frac{D_p \partial p / \partial x - D_n \partial n / \partial x}{p \mu_p + n \mu_n}
\end{equation}
with the Einstein relations $D_n=\mu_n k_B T / q$, $D_p=\mu_p k_B T / q$ for electrons and holes. As next to the anode contact 
\begin{equation}
\mu_pp \gg \mu_nn \text{~and~} -\mu_p \partial p / \partial x \gg \mu_n\partial n / \partial x 
\end{equation}
are valid assumptions, the electric field simplifies to
\begin{equation}
	F = \frac{k_BT}{q} \ln(p)
	\label{eqn:contactfield}
\end{equation} 
in analogy to organic bilayer solar cells.\cite{cheyns2008} Since the contact charge carrier density is the sum of thermally activated and accumulated charges by reduced surface recombination ($p - p_{th}$), this equation can be transformed to
\begin{equation}
 F = \frac{k_BT}{q} \left[ \ln(p_{th}) + \ln \left( 1 + \frac{p-p_{th}}{p_{th}} \right) \right].
\end{equation} 
The left summand is the electrical field of a sample with infinite surface recombination. By the right side the explicit effect of accumulated charges due to the majority surface recombination at the anode is considered. Therefore Eqn.~(\ref{eqn:Voc}) can be extended using the electrical field of a solar cell without blocking contacts $F_{\left( s \rightarrow \infty\right)}(x)$ and the separate charge carrier accumulation by the majority surface recombination velocity at the anode contact. Since at the cathode no additional holes are accumulated, we retrieve

\begin{equation}
	\begin{split}
	V_{oc} = \frac{E_g - \Phi_n - \Phi_p}{q} - \int_0^{L}  F_{\left( s \rightarrow \infty \right) }(x) \text{d}x\\
	- \frac{k_BT}{q}\ln\left(1+\frac{J_{s}}{q S^a_pp_{th}}\right)
	\end{split}
	\label{eqn:finalvoc}
\end{equation}
with a constant majority surface recombination velocity $S^a_p$, a required surface recombination current $J_s$ by minority charges and the thermally activated holes $p_{th}$ at the anode. An also reduced minority surface recombination can limit $J_s$ and therefore prevent a decrease of $V_{oc}$. By adding a PEDOT:PSS layer on the ITO substrate this effect was also seen experimentally.

Beside of surface parameters, $V_{oc}$ is influenced by bulk recombination processes (Eqn.~(\ref{eqn:langevin})), hence the charge carrier mobility.\cite{shuttle2008a, deibel2010review2} In Figure~\ref{fig:Voc}~b) the $V_{oc}$ dependences on balanced charge carrier mobilities at $10^{-10}~\text{m}^{2}/\text{Vs}$, $10^{-8}~\text{m}^{2}/\text{Vs}$ and $10^{-6}~\text{m}^{2}~\text{Vs}$ at fixed injection barriers of 0.1~eV and 0.3~eV are shown. High charge carrier mobilities possess a higher bulk recombination rate and therefore reduce $V_{oc}$.

\subsection{Space charge limited current} \label{sec:conductivity}

\begin{figure}
	\centering\includegraphics[width=9.0cm]{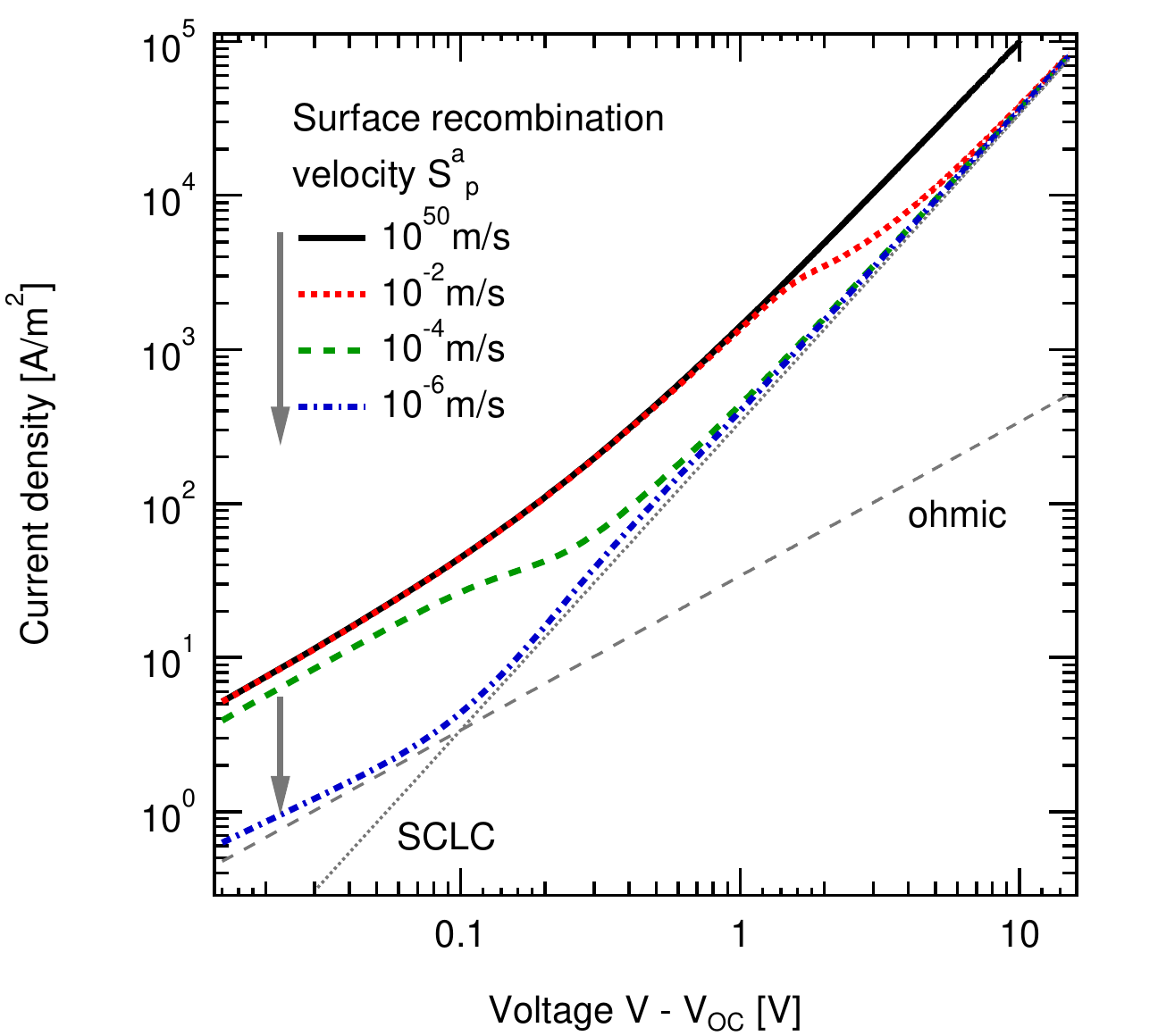}
	\caption{(Color online) Calculated current--voltage characteristics with $V_{oc}$ set to voltage axis origin. The anode majority surface recombination velocity is reduced from infinity to $10^{-6}$~m/s. At voltages near $V_{oc}$ an ohmic transport regime (slope 1) can be found, which changes to a space charge limited current at higher voltages (slope 2). The transition between both transport regimes is indicated by an s-shaped deformation of the calculated characteristics.}
	\label{fig:IVloglog}
\end{figure}

In order to extract information on the type of charge transport it is appropriate to set $V_{oc}$ as origin of the voltage axis. Subtracting $V_{oc}$ as seen in the lower right section of Figure~\ref{fig:IVlinear} from the voltage axis and choosing a double logarithmic presentation for the calculated current--voltage curves (Figure~\ref{fig:IVloglog}) reveals an ohmic ($J \propto V$) behavior at low voltages. At higher voltages it asymptotically changes towards a trap free space charge limited current (SCLC) defined as 
\begin{equation}
 J_{SCLC}= \frac{9}{8} \mu \epsilon_0 \epsilon_r \frac{V^2}{L^3}.
  \label{eqn:SCLC}
\end{equation}
This kind of charge transport was intensively studied in the past,\cite{pope1999, bohnenbuck2006, mihailetchi2005, arkhipov2001, jain2001} although the origin of s-shaped current--voltage kinks was not described.

An s-shaped deformation created by a finite majority surface recombination velocity connects this behavior to the analytic SCLC prediction. Hence, in context of Fig.~\ref{fig:IVloglog}, an s-shape can be interpreted as a direct transition from an ohmic toward a space charge limited current.

\section{Conclusion}

Using an extended oxygen plasma etching process of the transparent ITO anode during organic solar cell processing, we were able to generate s-shaped current--voltage characteristics. Due to the absence of planar semiconductor heterojunctions we were able to exclude local potential wells as origin of the observed kink. Nevertheless, assuming finite surface recombination velocities this behavior was qualitatively reproduced in a numeric macroscopic device simulation.

By analyzing the calculated energetic band structure, we assign space charges created by the reduced majority surface recombination velocity as origin of the characteristic deformation. Since also injection barriers change the charge carrier density at the contacts, we could see both parameters affecting $V_{oc}$, but only the surface recombination velocity being able to create an s-shaped current--voltage characteristic. Also in multilayer devices local space charges are responsible for the s-shape generation.

We presented an analytic approximation showing that the loss of open circuit voltage on decreased majority surface recombination velocities can be countervailed by the reduction of the minority surface recombination velocity. Finally we were able to show the s-shape indicating the transit from an ohmic conductivity to a space charge limited current.

\section*{Acknowledgment}
The authors would like to thank the German Federal Ministry of Education and Research (BMBF) for financial support in the frameworks of the MOPS project (contract no. 13N9867) and the OPV Stability project (contract No. 03SF0334F). C.D. acknowledges the support of the Bavarian Academy of Science and Humanities. V.D's work at the ZAE Bayern is financed by the Bavarian Ministry of Economic Affairs, Infrastructure, Transport and Technology.
 

\end{document}